\documentclass[12pt,preprint]{aastex}

\begin{document}

\title{Hierarchical Star Formation in the Spiral Galaxy NGC 628}
\author{Bruce G. Elmegreen\altaffilmark{1},
Debra Meloy Elmegreen\altaffilmark{2}, Rupali
Chandar\altaffilmark{3}, Brad Whitmore\altaffilmark{4}, and
Michael Regan\altaffilmark{4} \altaffiltext{1}{IBM Research
Division, T.J. Watson Research Center, 1101 Kitchawan Road, Yorktown
Heights, NY 10598; bge@watson.ibm.com} \altaffiltext{2}{Department
of Physics \& Astronomy, Vassar College, Poughkeepsie, NY 12604;
elmegreen@vassar.edu} \altaffiltext{3}{Center for Astrophysical
Sciences, Department of Physics \& Astronomy, The Johns Hopkins
University, Baltimore, MD 21218; rupali@pha.jhu.edu}
\altaffiltext{4}{Space Telescope Science Institute, 2700 San
Martin Drive, Baltimore, MD 21218; whitmore@stsci.edu;
mregan@stsci.edu}
 }

\begin{abstract}
The distributions of size and luminosity for star-forming regions
in the nearby spiral galaxy NGC 628 are studied over a wide range
of scales using progressively blurred versions of an image from
the Hubble Space Telescope Advanced Camera for Surveys. Four
optical filters are considered for the central region, including
H$\alpha$.  Two filters are used for an outer region. The features
in each blurred image are counted and measured using {\it
SExtractor}. The cumulative size distribution is found to be a
power law in all passbands with a slope of $\sim{-1.5}$ over 1.8
orders of magnitudes. The luminosity distribution is approximately
a power law too, with a slope of $\sim-1$ for logarithmic
intervals of luminosity.  The results suggest a scale-free nature
for stellar aggregates in a galaxy disk. Fractal models of thin
disks reproduce the projected size distribution and suggest a
projected mass distribution slope of $\sim-0.5$ for these extended
regions. This mass slope converts to the observed luminosity slope
if we account for luminosity evolution and longer lifetimes in
larger regions.
\end{abstract}

\section{Introduction}
Star formation by gravitational collapse and turbulence
compression leads to a hierarchy of structures ranging from star
complexes on a galactic scale to OB associations and subgroups
inside them, down to clumps of protostars inside embedded clusters
(see reviews in Efremov 1995; Elmegreen et al. 2000; Elmegreen
2002; Elmegreen 2005).

The largest several levels of this structure have been studied for
the LMC (Feitzinger \& Braunsfurth 1984; Feitzinger, \& Galinski
1987; Maragoudaki et al. 1998; Harris \& Zaritsky 1999), SMC (Chen
\& Crone 2005), M31 (Battinelli, Efremov \& Magnier 1996), NGC 300
(Pietrzyn'ski et al. 2001), M51 (Bastian et al. 2005), M33 (Ivanov
2005) and for several other galaxies in larger surveys (Bresolin
et al. 1998; Gusev 2002; Elmegreen \& Elmegreen 2001; Crone 2005).
A power-law autocorrelation function for clusters in the Antennae
galaxies suggests the hierarchy extends up to 1 kpc (Zhang, Fall,
\& Whitmore 2001). Power-law power spectra of optical light in
several galaxies suggest the same maximum scale, possibly
including the ambient galactic Jeans length (Elmegreen, Elmegreen,
\& Leitner 2003; Elmegreen et al. 2003).  If the ambient Jeans
length is the largest scale, then a combination of gravitational
fragmentation and turbulent fragmentation, with some of the
turbulent energy coming from gravitational self-binding, could
drive the whole process.  Krumholz \& McKee (2005) show how the
observed star formation rates in galaxies can follow from such
wide-ranging turbulent structures.

Very young clusters have a similar pattern of sub-clustering, but
on much smaller scales. For example, pre-main sequence stars in
Taurus have a power law 2-point correlation function, indicating a
wide range of scales for sub-clustering (Gomez et al. 1993). Also
found to be hierarchical are the T Tauri stars in NGC 2264 (Dahm
\& Simon 2005), the mm-wave continuum sources in the core of rho
Ophiuchus (M. Smith, et al. 2005), the protostars in the Serpens
core (Testi et al. 2000), massive star formation in the LMC OB
association LH 5 (Heydari-Malayeri et al. 2001), and the embedded
clusters in the W51 giant molecular cloud (Nanda Kumar, Kamath, \&
Davis, 2004).  Hierarchical triggering of star formation was noted
in the W3/4/5 region by Oey et al. (2005). Very small stellar
systems, consisting of only three or four stars inside a cluster,
can also be hierarchical (e.g., Brandeker, Jayawardhana \& Najita
2003), suggesting this structure continues down to individual
stars. An analogous continuation of massive stellar clustering
down to individual O stars was noted for the LMC by Oey, King, \&
Parker (2004).

Hierarchical clustering disappears with age as the stars mix. The
densest regions have the shortest mixing times and lose their
substructures first. The largest regions are generally not
self-bound and eventually lose their substructures after random
initial motions and tidal forces separate the individual clumps.
In general, the boundary between a smoothly distributed star
cluster and the surrounding hierarchical star field composed of
other clusters, OB associations, and star complexes, depends on
the density profile, the age, and the critical tidal density from
the surrounding galaxy. If the stellar density is higher than the
tidal density, then regions that are older than a crossing time
should be mixed and regions that are younger should still be
hierarchical. For example, in the Orion nebula, the youngest
objects, the proplyds, still cluster near $\theta^1$ Ori C while
the slightly older stars are more dispersed (N. Smith et al. 2005;
Bally et al. 2005). Simulations of collapsing clouds by Bonnell,
Bate \& Vine (2003) also show hierarchical star formation with
well-mixed clusters in several dense cores.

The purpose of this paper is to study hierarchical structure in
the SA(s)c galaxy NGC 628 using Hubble Space Telescope (HST)
observations of the central $\sim10$ kpc in 4 passbands and of an
outer region with the same size in 2 passbands. This extends a
similar study we did with HST images of ten other galaxies using a
slightly different technique (Elmegreen \& Elmegreen 2001).  In
both cases, objects were counted on progressively blurred images,
but here we consider the whole galaxy instead of individual star
complexes, and we also use {\it SExtractor} to enable counts in
the tens of thousands on pixel scales.  We also consider 4
passbands here to study the possible effects of extinction,
whereas before we used only one.  Models were compared with the
observations in both studies. Here we use models of star fields
based on the fractal Brownian motion technique and we apply
automated cloud counting methods to the results.  Before we built
fractals that were nested hierarchically and counted by eye. The
current technique allows more control over the power spectrum and
density probability distribution function, which are matched to
observations and turbulence simulations.

The order of this paper is as follows. Section 2 describes the
observations, Section 3 has the data analysis and measurements,
and Section 4 presents the resultant size and luminosity
distributions for stellar groupings. In Section 5, we present
several different models of star fields that represent star
formation in a turbulent gas having a power-law power spectrum.
The model results are then compared with the measurements for NGC
628.  The conclusions are in Section 6.

\section{Observations}
\begin{figure}
\epsscale{1.0} \plotone{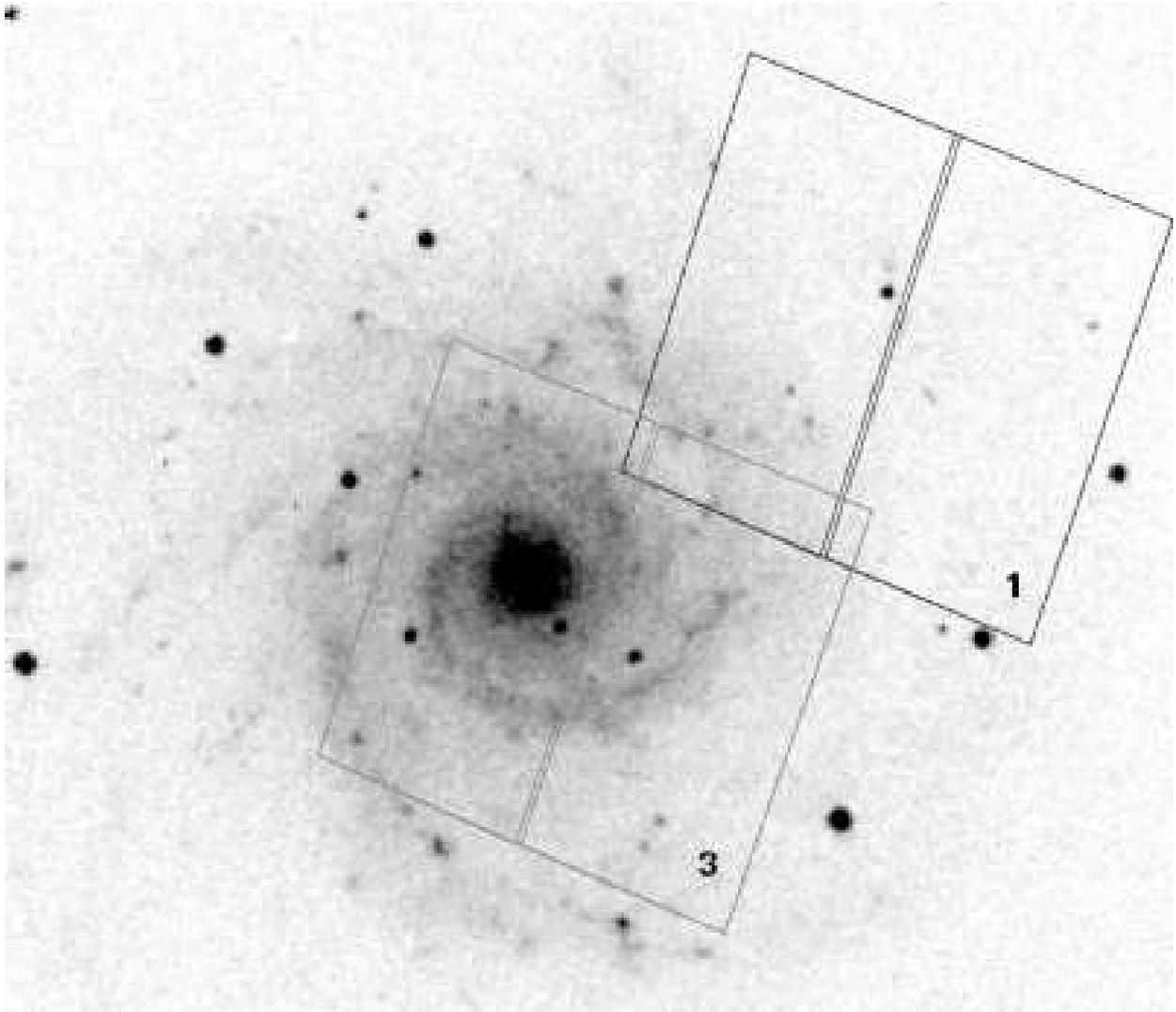} \caption{The Digital Sky
Survey B-band image of NGC 628 with ACS grids for positions 1
(Outer Region) and 3 (Central Region). North is up. The ACS boxes
are 210 arcsec across.}\label{N628ACS}\end{figure}

\begin{figure}
\epsscale{1.0} \plotone{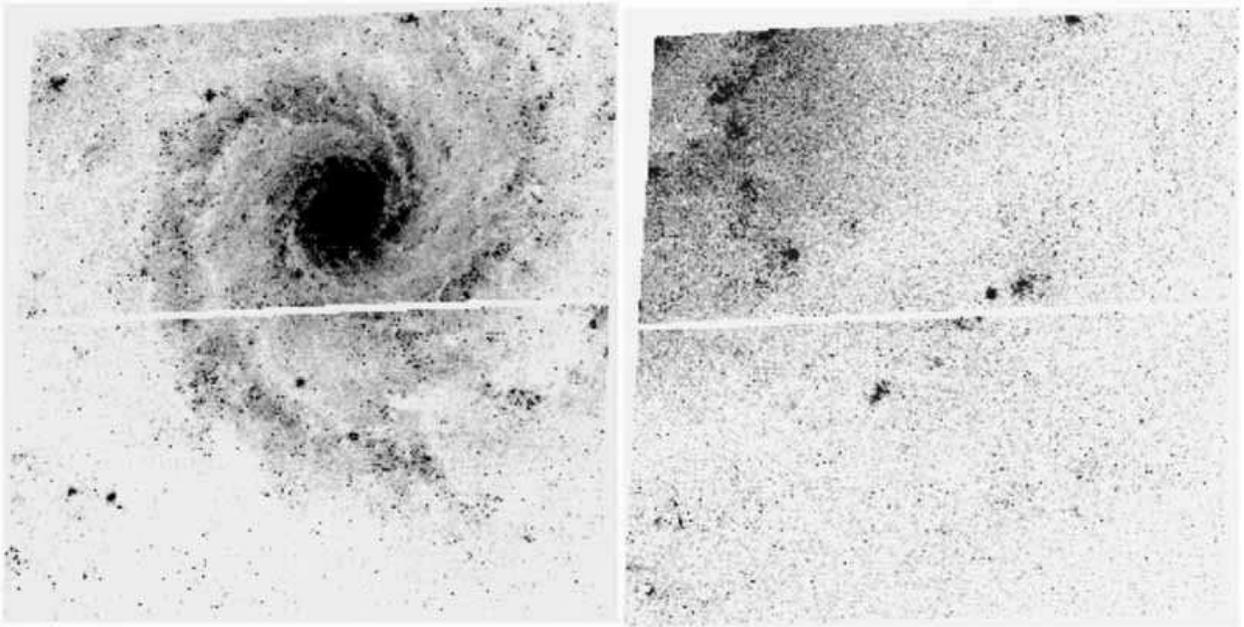} \caption{The ACS F435W
images of NGC 628. The orientation is 67$^\circ$ CCW from N in
both images. The images are 210 arcsec in
width.}\label{f1n628}\end{figure}

The HST Advanced Camera for Surveys Wide Field Camera (ACS WFC)
was used to image NGC 628 as part of a 37-orbit project on 7
nearby spiral galaxies to study star formation and structure (GO
10402; PI: Chandar). For the present study, we used images of NGC
628 centered on two different positions: a central one and an
outer one (as shown in Figure \ref{N628ACS}).  The central
position was observed with four filters: B (F435W), V (F555W), I
(F814W), and H$\alpha$ (F658N), while the outer position was
imaged in B and V. Figure \ref{f1n628} shows the B-band images for
the two positions, and Table 1 lists the positions and exposure
times. The H$\alpha$ image was continuum-subtracted by combining
the V and I images and scaling them so that the stars were
eliminated in H$\alpha$. The images are 4211x4237 pixels, with a
scale of 0.05 arcsec per px.

The distance to NGC 628 is assumed for simplicity to be 10 Mpc,
using the value in Tully (1988), who derived 9.7 Mpc.  Sohn \&
Davidge (1996) and Sharina et al. (1999) got $\sim7$ Mpc from on
red and blue supergiants.  For 10 Mpc, the pixel size of 0.05
arcsec equals 2.42 pc.

\section{Data Analysis}
The images were Gaussian blurred in IRAF (Image Reduction and
Analysis Facility) using the $\it{gauss}$ function with {\it
sigma} (radii) of 1, 2, 4, 8, 16, 32, and 64 pixels in order to
highlight features of different physical sizes.

The program $\it{SExtractor}$ was used to identify star-forming
regions on each image. The resultant source maps were compared
with the original images; the number of regions on the
Gaussian-blurred source maps was counted by eye for comparison
with what $\it{SExtractor}$ found, in order to check for proper
settings. Thresholds equal to $3\sigma$, $5\sigma$, and $10\sigma$
above the background were a reasonable range. Limits were set so
that the number of contiguous pixels above the threshold (the
parameter $\it{detect\_minarea}$ in $\it{SExtractor}$) was
required to be the square of the blur size. For example, for
2-pixel blurring, the objects had to have an area of at least 4
contiguous pixels; for 4-pixel blurring, the objects had to have
an area of at least 16 pixels, and so on. Saturated stars were not
included in the source counts.

\begin{figure}
\epsscale{1.0} \plotone{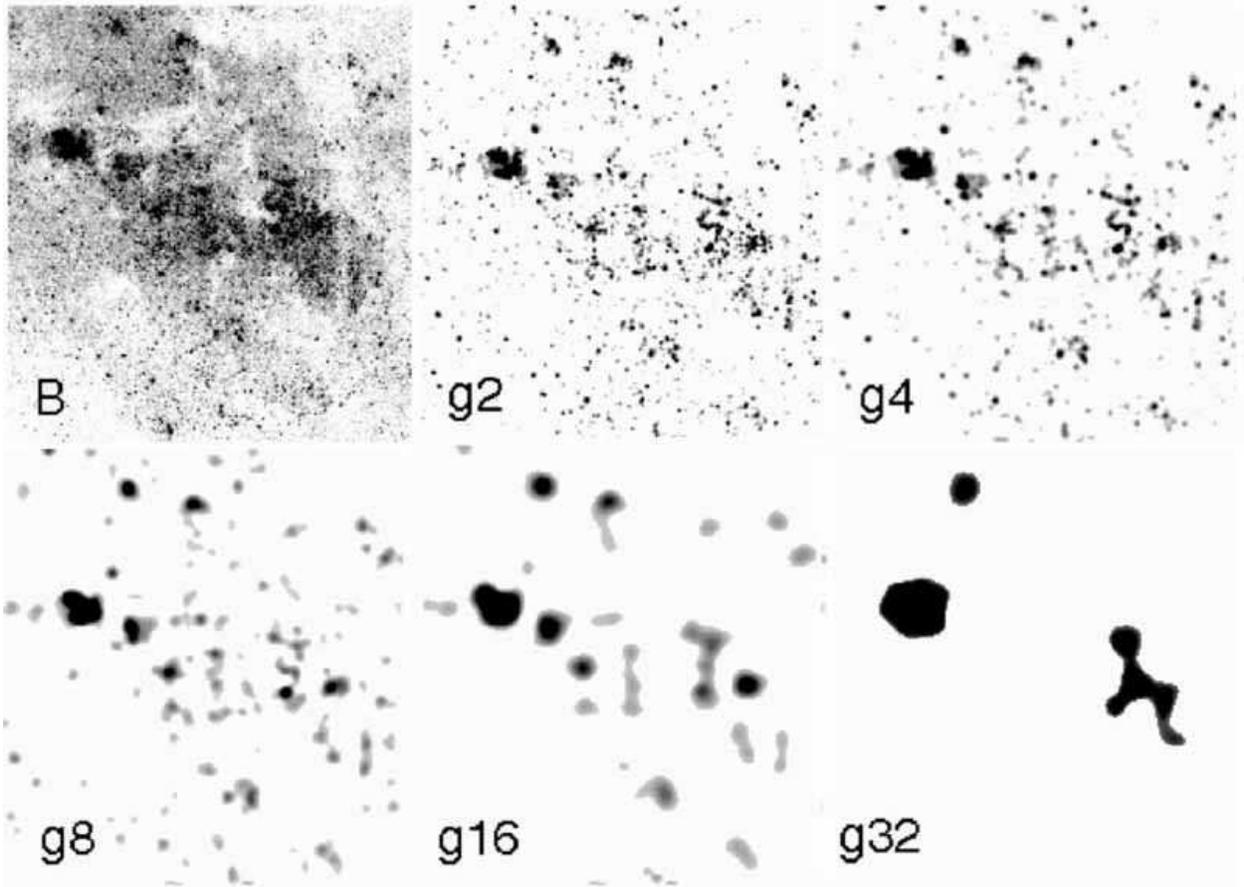} \caption{The F435W image
(labeled B) of a star-forming complex in the central region is
shown along with the $\it{SExtractor}$ object fields for
Gauss-blurred images. The level of blurring is indicated by the
number; g2 is a Gaussian blur of 2 pixels,
etc.)}\label{f2n628}\end{figure}

Figure \ref{f2n628} shows a B-band image of an enlarged portion of
the Central Region. Also shown are the object fields generated by
$\it{SExtractor}$ with a 5$\sigma$ threshold applied to images
that are Gauss-blurred with 2, 4, 8, 16, and 32 pixels.
$\it{SExtractor}$ finds all of the features that are {\it larger}
than the blurring size.

\section{Results}

\subsection{Size distribution function}
\label{size}

\begin{figure}
\epsscale{1.0} \plotone{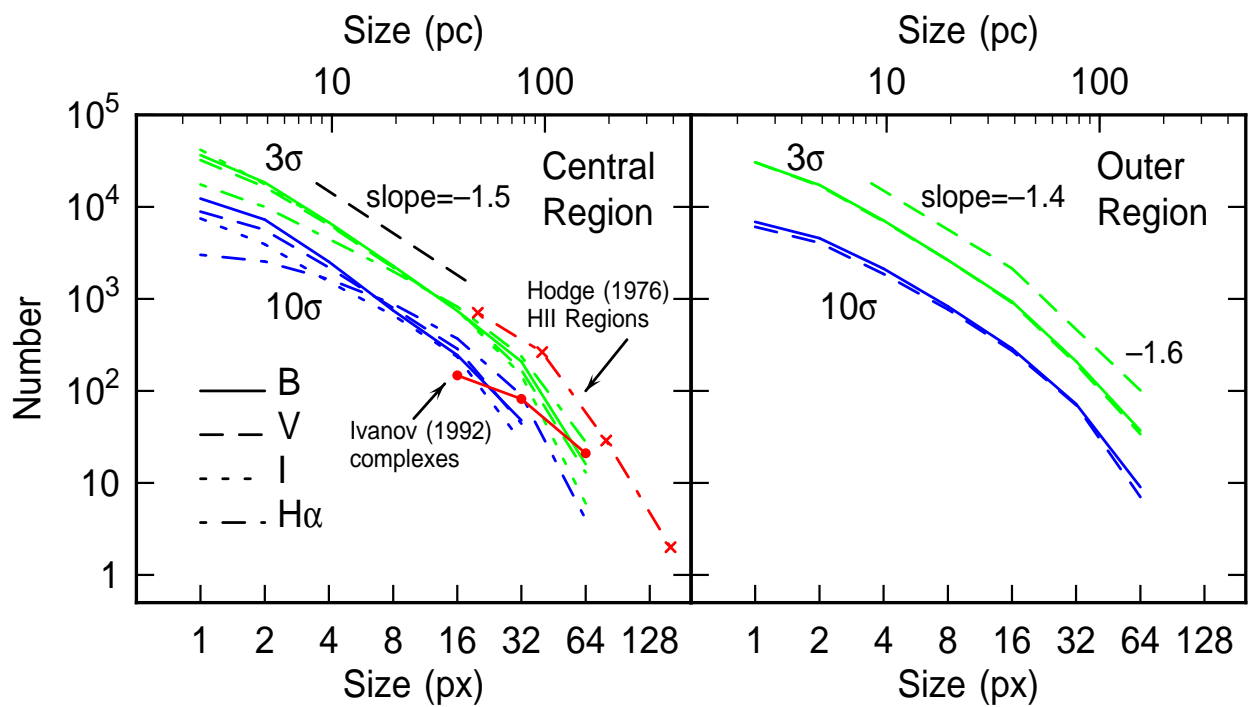} \caption{Cumulative size
distributions for regions brighter than the 3$\sigma$ or
10$\sigma$ noise limits and larger than a given Gaussian-blur
size. Different curves correspond to B,V,I, and H$\alpha$ images
as indicated. The central region of NGC 628 was used for data on
the left and the outer region was used for the right. Only B and V
images are available for the outer region. The size is on the
abscissa in units of pixels. The distributions are similar
regardless of noise limit, filter, and position; all show a power
law slope of $\sim -1.5$ with downward curvature at the upper end
that is presumably from image edge effects. The cumulative
distributions of size for HII regions and star complexes in this
galaxy are also shown in the left-hand panel, with size converted
to ACS pixels for comparison to the HST data.}\label{n628n-s}
\end{figure}

Figure \ref{n628n-s} shows the cumulative size distributions for
star-forming regions. The Gaussian blur radius in pixels is
plotted on the abscissa and the number of regions having radii
greater than the Gaussian blur radius is plotted on the ordinate.
The distributions are shown for the $3\sigma$ and $10\sigma$
limits in $\it{SExtractor}$ and for all of the available ACS
passbands for the Outer and Central Regions, respectively. Star
complexes and associations identified by eye on B-band
ground-based images by Ivanov et al. (1992) are also included in
this figure. The sizes given by Ivanov were divided by 2 to
convert to radii and summed with radius bins equal to those of our
ACS data. The smallest scale in the Ivanov compilation,
$1.6^{\prime\prime}$, is close to the seeing limit, so the
shallower slope of that distribution compared to ours may be the
result of incompleteness. HII region sizes measured by Hodge
(1976) on H$\alpha$ photographic plates of NGC 628 are included in
Figure \ref{n628n-s} too. The sizes were again divided by 2 and
summed using the bins given in Hodge's paper: $2^{\prime\prime}$,
$4^{\prime\prime}$, $8^{\prime\prime}$, and $16^{\prime\prime}$,
which correspond to 20, 40, 80, and 160 px in the ACS. The Hodge
distribution of HII regions matches fairly well the distribution
found here at the largest scales. Kennicutt \& Hodge (1976) found
from a power spectrum analysis of HII regions along the two main
spiral arms that there is no preferred clump size.

Figure \ref{n628n-s} indicates that the size distributions for
emission features are slightly curved downward, as would be the
case for a log-normal distribution, but resolution limits at small
scales and spiral arms, disk thickness, or image edge effects at
large scales could produce this curvature also (see Sect.
\ref{models}).

There is little difference between the $3\sigma$ and $10\sigma$
features measured by $\it{SExtractor}$. This cutoff similarity,
and the similarity in slope of the distribution functions for the
different passbands, suggests that what we are measuring is the
emission from hierarchically clustered stars, not the emission
from a smooth disk viewed through hierarchically clustered dust
holes. Dust holes should not give high brightness contrasts for
all hole sizes like what we observe with the $10\sigma$ cutoff.
The $3\sigma$ curves are higher than the $10\sigma$ curves because
there are about 3 times as many features at $3\sigma$ above the
noise than there are at $10\sigma$ above the noise. This means
that the size distribution of the bright stellar cores (i.e.,
those $10\sigma$ above the noise) has about the same slope as the
size distribution of the faint stellar envelopes (those $3\sigma$
above the noise). This similarity is consistent with the overall
hierarchical nature of the star fields themselves, as shown by the
near-power law size distributions. Generally, we expect the size
distribution to get steeper as the cutoff brightness is increased
because the objects selected at high cutoff levels are more and
more limited to single emission points as a result of resolution
limits and individual stars. This trend will be evident in the
model below (see Fig. \ref{fig:c1}). The curves in Figure
\ref{n628n-s} are apparently not at such high cutoff levels.

The B, V, and I band size distributions are close to power laws
all the way down to the smallest size, which corresponds to the
angular resolution of the original image, about 1 pixel in radius.
There is a slight drop at the smallest scale for the V-band image
at $10\sigma$ in the Central Region, but no significant drop at
the smallest scale for the other bands in this region. This is in
contrast to the Outer Region, which has distinct deviations from
power laws at the smallest scales in both B and V.

There are also deviations from power laws for the HII region size
distributions at the smallest scale. The HII region distribution
was measured only for the galaxy Central Region because the
continuum subtraction for the H$\alpha$ image was made from a
weighted average of the V and I bands, and there was no I band
exposure for the outer region. The drop at small size for HII
regions is much larger than the drop at small sizes for stellar
structures in the V-band image at $10\sigma$. There is a factor of
$\sim10$ fewer HII regions with a size comparable to the image
resolution than would be expected from an extrapolation of the
power laws at larger scales. Recall that 1 and 2 pixels correspond
to 2.7 and 5.4 pc at the assumed distance. Thus the drop suggests
there are relatively few tiny HII regions. This result would be
expected if most of the small HII regions represent only a
short-lived phase in the expansion of a nebula around a massive
star.

The slope of the power law for the size distribution is about
$-1.5$, as suggested by the dashed lines in the figure.  For a
fractal of dimension $D$, the integrated number of objects having
a size $\Sigma$ greater than some size $S$ scales as
$N_\Sigma(\Sigma>S)\propto S^{-D}$.  The number having a size
between $S$ and $S+dS$ scales as $n_S(S)dS\propto S^{-1-D}dS$. The
first function is the integral over the second,
$N_\Sigma(\Sigma>S)=\int_{\Sigma}^{\infty} n_S(S)dS$.  Taken at
face value, the size distribution of stellar aggregates suggests a
fractal distribution of stellar positions projected on the disk of
the galaxy having a fractal dimension of $D=1.5$.  This is less
than 2, as expected, because the stellar aggregates do not
completely fill the 2-dimensional plane of the galaxy disk. It is
comparable to the fractal dimension of projected local
interstellar clouds, which is $\sim1.3$ (e.g., Falgarone et al.
1991), to the fractal dimension of HI ($D=1.2-1.5$) in the M81
group galaxies (Westpfahl et al. 1999), and to the fractal
dimension of young stars in several dwarf galaxies (Crone 2005).
Evidently, stars form in a fractal gas and preserve that pattern
for some time.

\subsection{Luminosity distribution functions}

\begin{figure}
\epsscale{1.0} \plotone{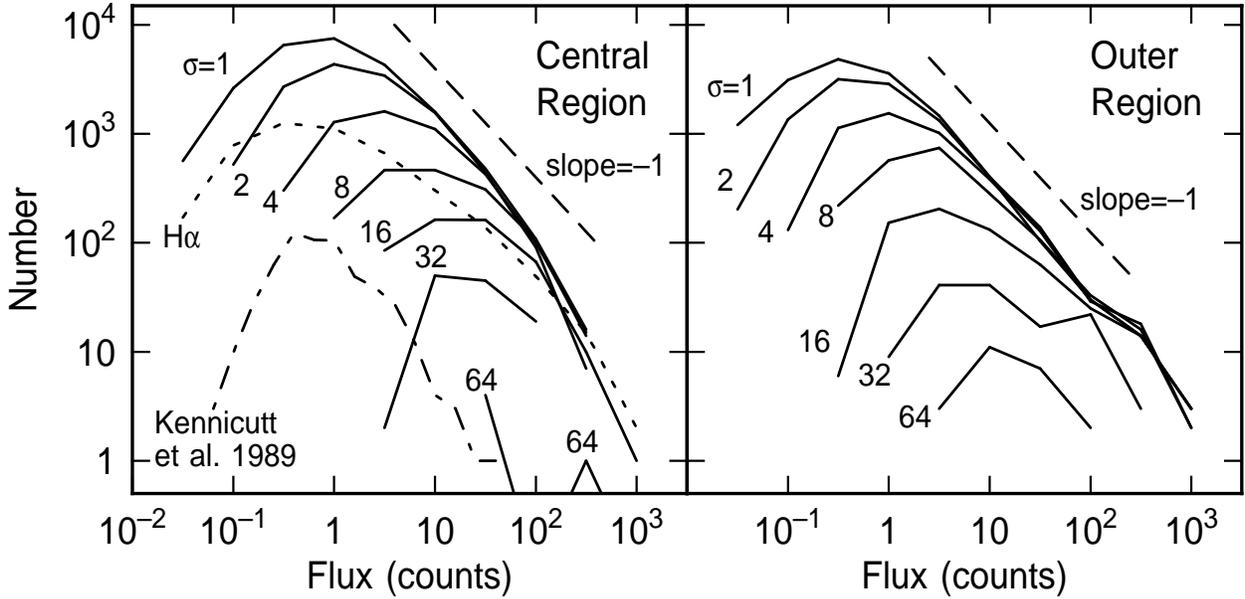} \caption{Number distributions of
flux count rates in logarithmic intervals for stellar aggregates
observed in B-band and H$\alpha$ (dotted line) ACS images of the
central region (left), and for the B-band image of the outer
region (right). Each solid curve is for a different Gaussian blur
size limit in B-band, as indicated by the associated numbers of
pixels for the blur. For the H$\alpha$ curve, only the 2-pixel
Gaussian blur distribution is shown (the others have a nested
structure similar to that of the B band). The slopes at high flux
are all similar and about equal to $-1$ as indicated by the solid
line. The curves turn over at the faint end because of the
Gaussian blur. Flux is defined to be the sum of the count rates
for the pixels in each region, in units of electrons per second. A
count rate of 1 for the B band corresponds to an apparent
magnitude of 25.15; a count rate of 1 in H$\alpha$ corresponds to
$1.6\times10^{38}$ erg s$^{-1}$.}\label{n628lf}\end{figure}

Figure \ref{n628lf} shows the luminosity functions of the stellar
concentrations found in each Gaussian-blurred image. The ordinate
is the count of regions having luminosities within a logarithmic
interval of 0.5 on the abscissa.  The abscissa is the flux of the
various regions measured in total calibrated counts s$^{-1}$ on
the ACS image. The calibration is such that a rate of 1 count
s$^{-1}$ corresponds to an apparent magnitude of 25.15 mag using
conversions in the on-line ACS data handbook. The blurring radius
is indicated near each curve. The solid curves are for the B-band
image with a $5\sigma$ lower limit in $\it{SExtractor}$. The
curves nest inside each other at the high luminosity side because
they all include the same regions at this end. The smaller regions
become progressively more prominent as the resolution improves,
filling out the lower luminosity end. The HII region luminosity
function is shown as a dotted curve, using the 2-pixel Gaussian
blurred image with a $5\sigma$ detection limit. It is
significantly shallower than the stellar curves. The HII region
luminosity function from Kennicutt, Edgar, \& Hodge (1989) is also
shown in Figure \ref{n628lf} as a dot-dash line. It was scaled to
the flux count rate for the H$\alpha$ ACS image and is similar to
the H$\alpha$ function found here but with lower count rate. The
calibration conversion for H$\alpha$ is that one ACS count
s$^{-1}$ corresponds to $1.6\times10^{38}$ erg s$^{-1}$ at 10 Mpc.

The fraction of the total B-band emission in the regions counted
in Figure \ref{n628lf} decreases with increasing blur radius. For
blurs of 1, 2, 4, 8, 16, 32 and 64 pixels, this fraction is 0.16,
0.13, 0.10, 0.07, 0.04, 0.01, and 0.0003. This is consistent with
the decreasing integrals under the curves in Figure \ref{n628lf}.
The decrease confirms our impression from Figure \ref{f2n628} that
the regions are not strictly hierarchically nested with all the
small regions inside the large regions. If this were the case,
then every region would be present in all blur images, with the
small region fluxes contributing to the large regions at large
blur radius; as a result, the fractional flux would be constant
with blur radius. Instead, there are numerous small regions
independent of the large regions. These small regions get lost at
large blur radius and drop below our detection cutoff, causing the
integrated flux to decrease with blur radius.  We note that our
models in Section \ref{models} look the same as the observations
in the sense that there are numerous small regions independent of
the large regions, in addition to small regions embedded inside
the large regions.

The slope of the luminosity distribution function for stellar
aggregates is $\sim-1$ on this log-log histogram. This means the
luminosity distribution is approximately $N_{log}(L)d\log L
\propto L^{-1}d\log L$, or in linear intervals, $n_L(L)dL\propto
L^{-2}dL$.

This slope is typical for the luminosity distributions of standard
(``open'') clusters, as found in many studies.  Kennicutt \& Hodge
(1980) found a similar distribution for HII region luminosities in
NGC 628 from ground-based images, which did not include the
central 1$^{\prime}$ of the galaxy and included only nebulae
larger than about 4$^{\prime\prime}$.  Lelievre \& Roy (2000)
found a slope of $\sim{-1.6}$ on a cumulative luminosity function,
which implies the same slope of $-1.6$ on a histogram with
logarithmic binning, as in our Fig. \ref{n628lf}.

The slope of the luminosity function for extended star-forming
regions should not be viewed as a continuation of the slope of the
luminosity functions for young clusters and HII regions, even
though the numerical values of these slopes are about the same.
The extended distribution found here has a much larger scale than
any individual cluster or exciting source of an HII region, and
the larger region is likely to be significantly evolved, as for a
typical star complex (Efremov 1995).  Thus, the luminosity of an
extended region is not directly proportional to its mass, as it is
for a young cluster. The large regions are also likely to be a
composite of several smaller regions observed in projection
through the galaxy, unlike individual clusters and HII regions. We
return to these points in section \ref{models} where fractal
models reproduce the observed distributions.

The slope of the luminosity function, combined with the slope of
the size distribution, gives the relation between luminosity and
size. Taking $n_S(S)dS\propto S^{-2.5}dS$ and $n_L(L)dL\propto
L^{-2}dL$, and assuming a one-to-one correspondence where
$n_S(S)dS=n_L(L)dL$, gives $S^{-2.5}dS\propto L^{-2}dL$, from
which we derive $L\propto S^{1.5}$.  This power is another
definition of the fractal dimension because a fractal has its
content (mass, luminosity, number of objects, etc.) increase with
size to a power that is the fractal dimension, $D=1.5$ in this
case. The luminosity-size relation implies that the projected
stellar luminosity density decreases for larger regions as
$LS^{-2}\propto S^{-0.5}$. This is a sensible result because
larger stellar aggregates have more open space.

\section{Models}
\label{models}

\begin{figure}
\epsscale{0.7} \plotone{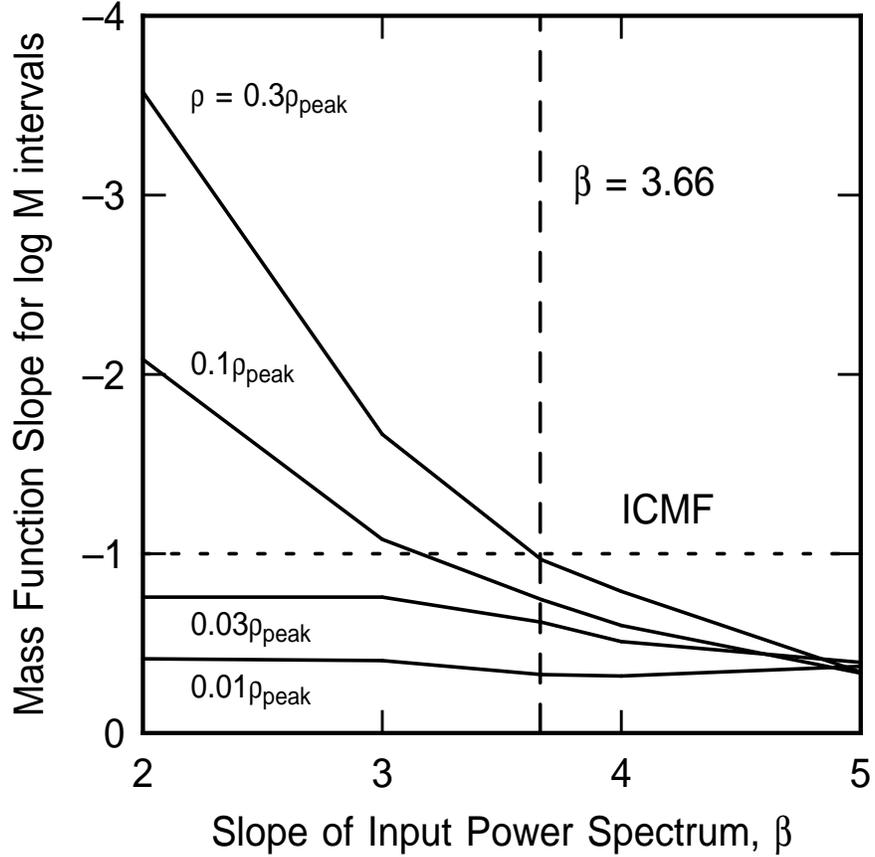} \caption{Slopes of the mass
distribution functions for clouds in three dimensional fractal
Brownian motion models. The abscissa is the slope of the noise
power spectrum used to generate the models.  The ordinate is the
slope of the cloud number distribution calculated for equal bins
in the log of the mass. The power spectrum slope for 3D Kolmogorov
turbulence is indicated by the vertical dashed line. The typical
slope of the initial cluster mass function (ICMF) is indicated by
a horizontal dotted line. Different curves are for different
density cutoffs in the definition of a cloud. Low density cutoffs
give a shallower slope, $\sim-0.5$, which is similar to the
observed slope of the molecular cloud mass distribution.  High
density cutoffs give steeper slopes, like the
ICMF.}\label{massslope}\end{figure}

Three-dimensional fractal models of density distributions having
an approximately power-law power spectrum like the ISM (e.g.,
Dickey et al. 2001) and a log-normal density probability
distribution function, as in turbulence simulations
(V\'azquez-Semadeni 1994; Elmegreen \& Scalo 2004), can reproduce
the observed $M^{-1}d\log M$ distribution of stellar clusters and
the more shallow distribution of molecular clouds (Elmegreen 2002;
2004). The shallow distribution comes from the same fractal as the
steep distribution, but with a lower density cutoff (see also
Sanchez, Alfaro \& P\'erez 2005). Figure \ref{massslope}
(reproduced from Elmegreen 2004) shows this result by plotting the
slope of the clump mass function versus the power $\beta$ of the
power spectrum, generated from the average of many random fractal
models, for various lower limits to the density measured in terms
of the peak density in the fractal. For a 3D Kolmogorov-type
spectrum, with $\beta=-11/3$ (vertical dashed line), the dense
cores have an $M^{-1}\log M$ distribution like star clusters and
the low density regions have shallower distributions, $\sim
M^{-0.5}d\log M$, like molecular clouds (e.g., Solomon et al.
1987).  Thus a fractal, log-normal ISM, such as that generated in
turbulence simulations, can reproduce both the GMC and the cluster
mass functions.

Our observations here show similar power-law distributions, but
the measured quantities are different. First, the observations
consider fairly large regions (the 64 px blur corresponds to
objects larger than 155 pc) and not individual clusters. Second,
these regions are observed in projection through a galaxy disk and
are probably not spherical anyway. To simulate this case, we make
a fractal Brownian motion model for the density distribution in a
face-on slab with a thin Gaussian profile on the line of sight.
This is made by first filling one-half of a $210^3$ cube in
wavenumber space ($k_x,k_y,k_z$) with random complex numbers
uniformly distributed from 0 to 1, and then multiplying each
number by $k^{-\beta/2}$ for $k^2=k_x^2+k_y^2+k_z^2$. The inverse
Fourier transform of this distribution is a multifractal density
cube in real space with a power spectrum slope $-\beta$. It has a
Gaussian density probability distribution function (pdf), so we
exponentiate it to give another density distribution which has a
log-normal density pdf. The power spectrum changes a little at low
$k$ after this exponentiation, but not noticeably at intermediate
to high $k$ (see Weinberg \& Cole 1992 for corrections to recover
the original power law). Our galaxy model uses the inner $180^3$
cube of this density distribution to avoid periodic edge effects,
and multiplies it by a Gaussian along the line-of-sight $z$ axis,
centered at the $z$ midplane and having a dispersion of 18 pixels.
Thus the model is a face-on multifractal slab (x,y plane) with a
Gaussian line-of-sight distribution for the mean density. It has a
ratio of width to effective depth equal to 5. In a final step, the
galaxy model was summed over the line-of-sight $z$ to make the
projected multifractal galaxy.

We are interested in the size and mass distributions of the
densest clouds in this multifractal galaxy model.  We define a
cloud as an isolated but internally connected region with all
pixel values above a certain cutoff. If the cutoff is too low,
then the regions are so large that they overlap and form only one
big cloud. If the cutoff is too high, then the regions are small
and there are only a few of them. We found that cutoff values in
the range between 0.3 and 0.6 times the peak density gave good
samples of clouds for the size and mass distribution functions.
The cloud boundaries are identified automatically by a random
walk. More details are in Elmegreen (2002). To determine
statistically significant distribution functions, we made 90
random models for each pair of parameters, $\beta$ and the cutoff
value, and then plotted the total count of clouds from all of
these models. Each model has a different peak density so the
cutoff varies a little from model to model in absolute terms for
each parameter pair.

\begin{figure}
\epsscale{1.0} \plotone{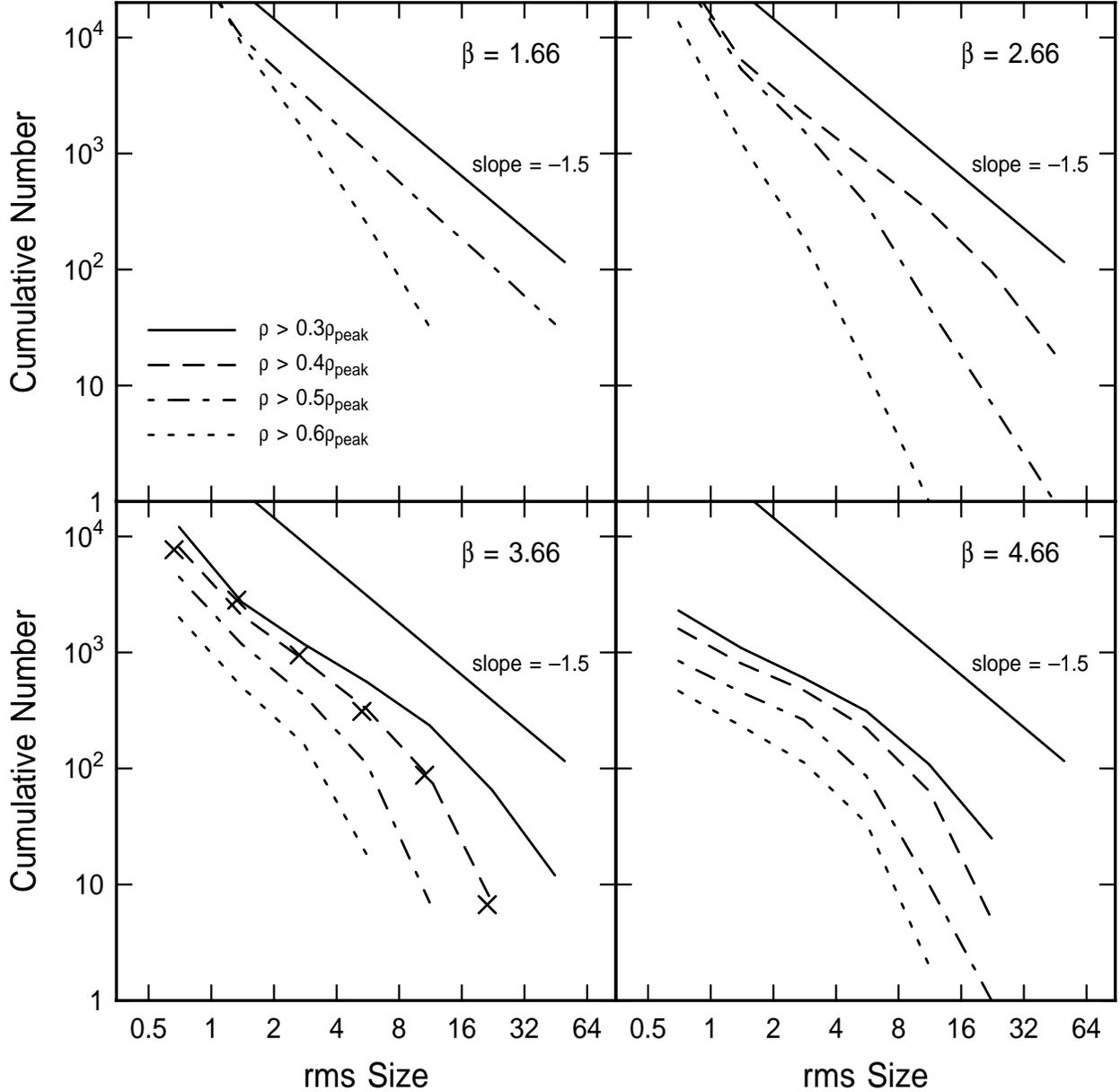} \caption{Cumulative projected size
distribution functions for clouds in fractal Brownian motion
models of face-on thin disks. The abscissa is the size of the
cloud in pixels and the ordinate is the number of clouds larger
than this size in the model. The model disks are projections of
multifractal density cubes with a Gaussian multiplier for the
average density on the line of sight. The aspect ratio of the disk
is 5:1. Each panel is for a different slope $\beta$ of the power
spectrum for the noise in the fractal Brownian motion model.
Different curves in each panel are for different density cutoffs
in the definition of a cloud. A fiducial slope of $-1.5$ is
indicated by a line. The crosses represent the observed size
distribution from Fig. 4 for B-band at the 3$\sigma$ limit.
}\label{fig:c1}\end{figure}

\begin{figure}
\epsscale{1.0} \plotone{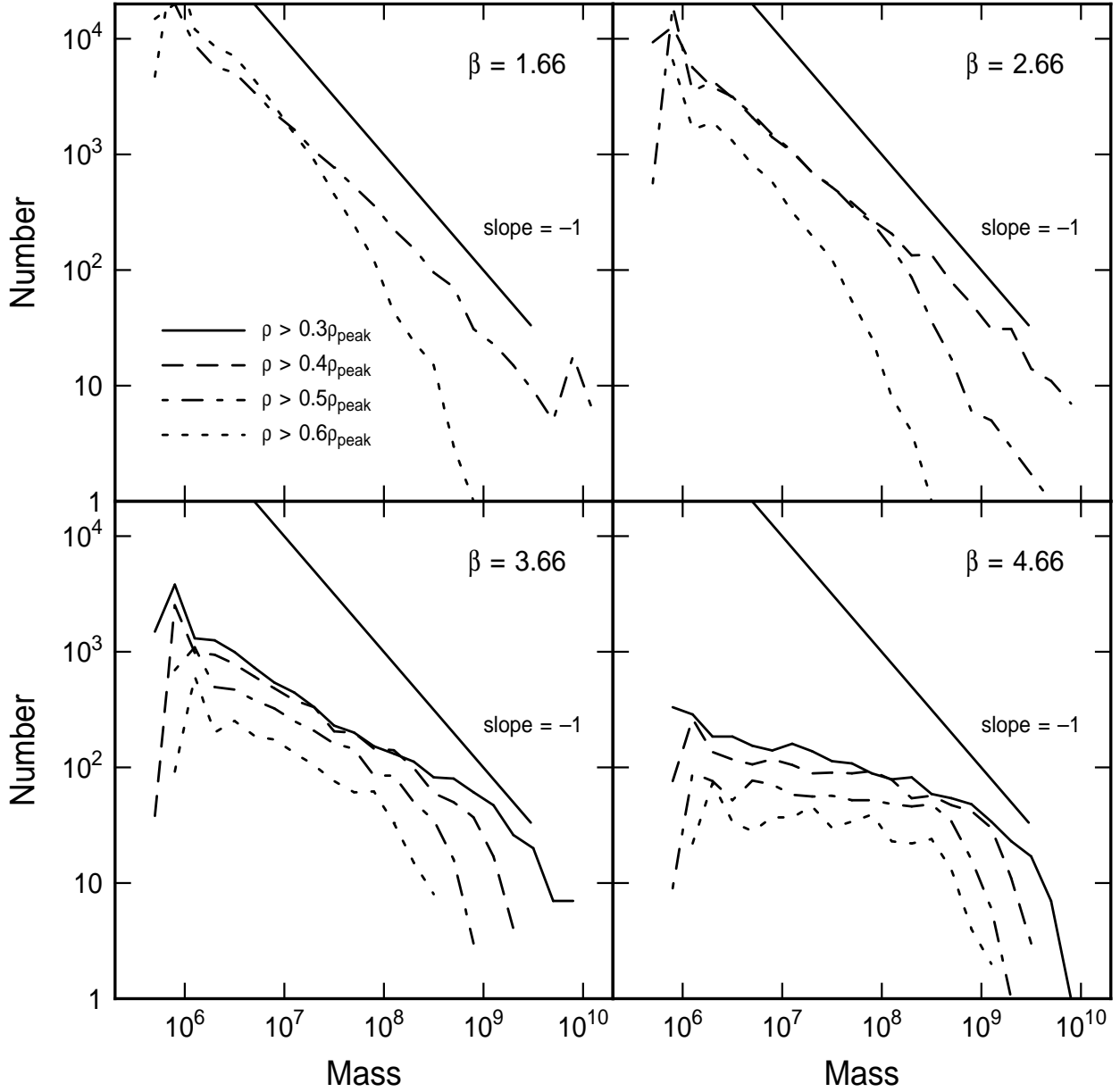} \caption{Mass distributions of
projected clouds in the fractal galaxy model. Each panel is a
different $\beta$, which is the slope of the power spectrum, and
each line in a panel is a different cutoff density for the
definition of a cloud. The low $\beta$ panels do not show the
results for low cutoff densities because in these cases most of
the gas blends into only a few big clouds. A fiducial slope of
$-1$ is shown.  }\label{fig:c2}\end{figure}

The cumulative size and mass distribution functions are shown in
Figures \ref{fig:c1} and \ref{fig:c2}.  Each figure has four
panels, one for each $\beta$, the slope of the power spectrum of
the multifractal. In each panel there are 2, 3, or 4 curves
corresponding to the four cutoff values. The low cutoff values for
low $\beta$ are not plotted because these clouds overlapped
severely.  The value of $\beta=3.66$ is appropriate for the
velocity power spectrum of three-dimensional Kolmogorov
turbulence; $\beta=2.66$ for two-dimensional Kolmogorov turbulence
as might be expected in a thin galaxy disk.  We show these values
for this reason plus the two other values on either side.  The
panel for $\beta=3.66$ has crosses representing the counts
observed in B-band for the $3\sigma$ limit (from the left hand
panel of Fig. \ref{n628n-s}).

The cumulative size distribution in Figure \ref{fig:c1} plots the
number of clouds larger than the size given on the abscissa, in
units of model pixels. The size of a cloud is defined to be the
density-weighted two-dimensional rms dispersion of all the
projected pixels in the cloud.  The rms size of a single pixel
cloud is taken to be 0.707 pixel units. The mass is defined to be
the sum of the densities in the cloud.

\begin{figure}
\epsscale{1.0} \plotone{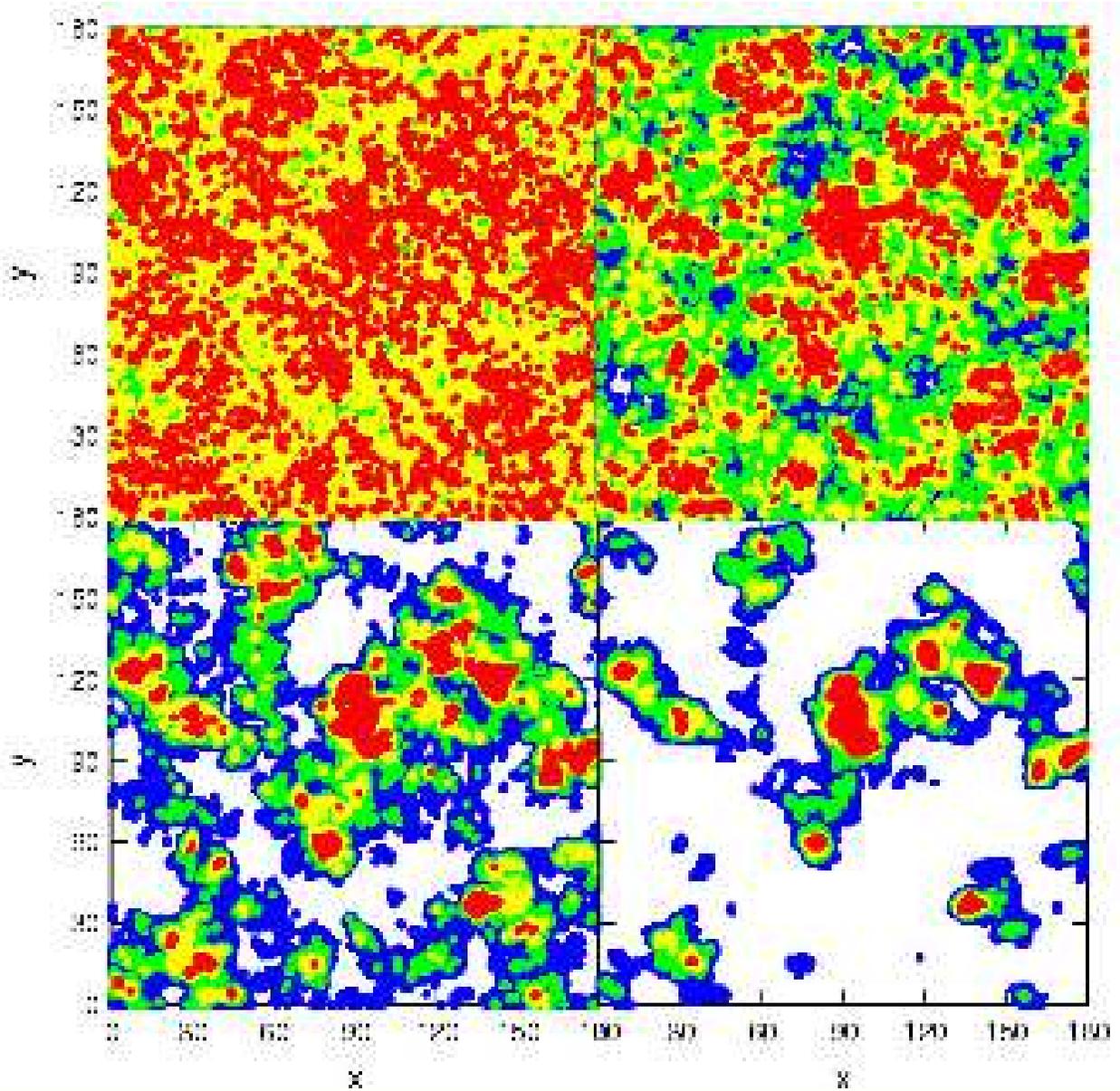} \caption{Images of projected
fractal galaxies for four different $\beta$ showing projected
density as color. The top left has $\beta=1.66$, top right
$\beta=2.66$, lower left $\beta=3.66$ and lower right
$\beta=4.66$. Blue regions have projected densities between 0.3
and 0.4 of the peak, green between 0.4 and 0.5, yellow between 0.5
and 0.6, and red greater than 0.6. (degraded resolution for
astroph) }\label{fig:fractal2}\end{figure}

Figure \ref{fig:fractal2} shows an example of each map for the
four different power spectrum slopes, $\beta$. The four cutoff
levels are indicated by color: blue pixels are between 0.3 and 0.4
of the peak, green between 0.4 and 0.5, yellow between 0.5 and
0.6, and red greater than 0.6. The order in $\beta$ of the panels
is the same as in the previous two figures. The random numbers are
the same for each panel, so the designs are similar; the only
difference is the value of $\beta$.

The cumulative size distributions in Figure \ref{fig:c1} curve
downward a little, with an average slope at mid-range of
approximately $-1.5$, as indicated by the solid line. The slopes
are systematically steeper for higher cutoff values, and the
counts are lower too. The counts also decrease slightly with
increasing $\beta$. The curvature at large size is from edge
effects: the rms size of the whole projected square is 73 px. The
mass functions in Figure \ref{fig:c2} are also somewhat curved
from edge or spiral arm effects but they have a power law middle
range with a slope between $-0.5$ and $-1$ for these value of
$\beta$.

The models agree fairly well with the observations (crosses in
Fig. \ref{fig:c1}), suggesting that the size distribution for
stellar aggregates in NGC 628 (Fig. \ref{n628n-s}) results from a
multifractal ISM where star formation occurs in the gas peaks. The
model size distributions are relatively insensitive to the power
spectrum, but the observed ISM slope $\beta\sim3$ (e.g.,
Stanimirovic et al. 1999; Dickey et al. 2001; Elmegreen, Kim, \&
Staveley-Smith 2001) gives about the right size distribution for
stellar aggregates.

\begin{figure}
\epsscale{.8} \plotone{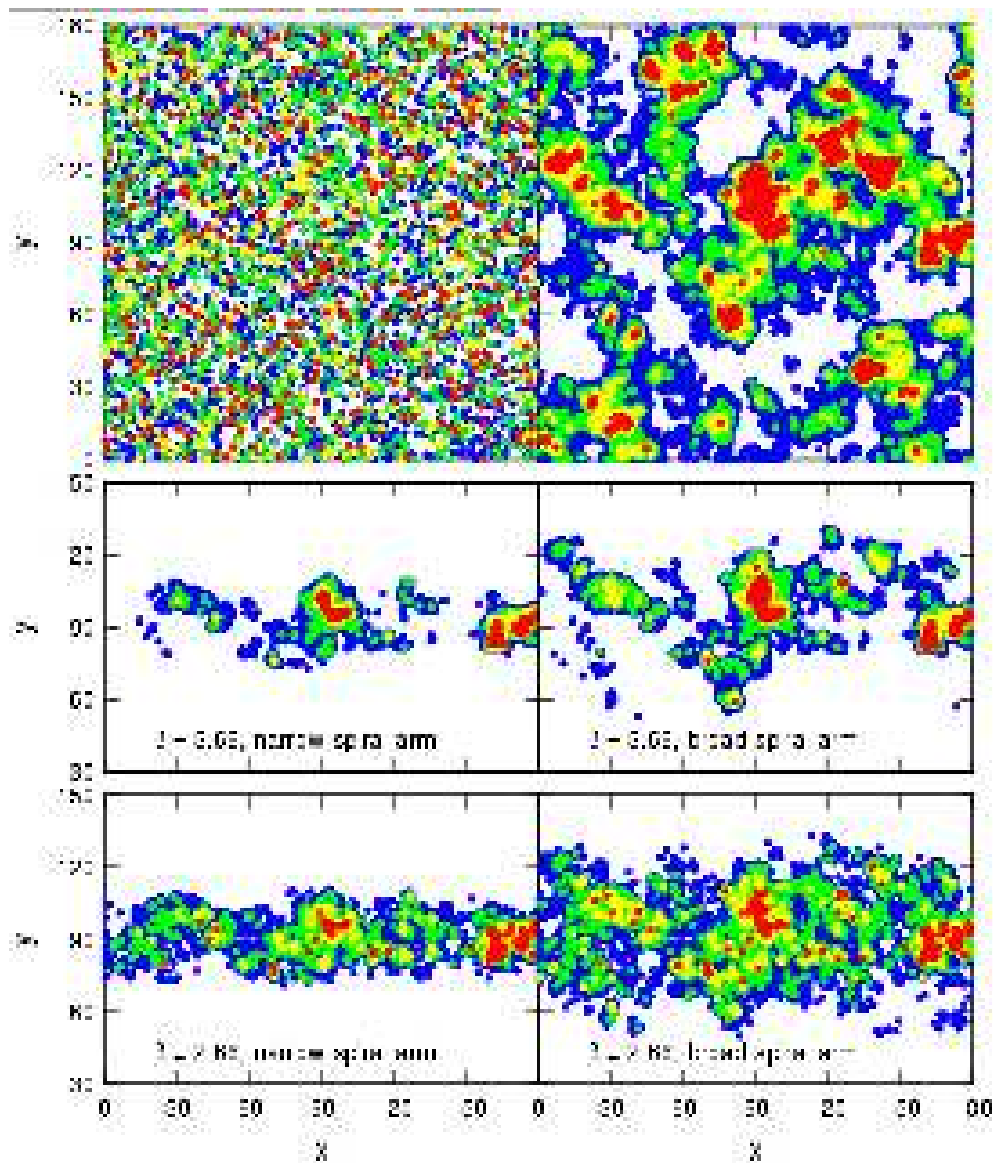} \caption{Images of
alternative models for projected fractal galaxies, all of which
have a Gaussian profile of average density on the line of sight
with a dispersion of 18 pixels. Top left: purely random values are
used for the density prior to multiplication by the line of sight
Gaussian. Top right: $\beta=3.66$ model reproduced from Fig. 9.
Middle left: $\beta=3.66$ model with a Gaussian profile in the $y$
direction having a dispersion equal to the dispersion of the line
of sight depth. Middle right: $\beta=3.66$ with a $y$ dispersion
twice as large. Bottom left and right: $\beta=2.66$ with the same
two $y$ dispersions. In the top two panels, the blue regions have
projected densities between 0.3 and 0.4 of the peak, green between
0.4 and 0.5, yellow between 0.5 and 0.6, and red greater than 0.6.
In the bottom four panels, the density limits are shifted because
the peak density is smaller on average when the models are
smaller: blue is 0.4-0.5 times the peak, green is 0.5-0.6, yellow
is 0.6-0.7, and red is greater than 0.7 times the peak. (degraded
resolution for astroph)}\label{fig:fractal7}\end{figure}

Other types of models were run for comparison. A model galaxy
where every pixel in real space was noise, rather than the Fourier
transform of noise with a power-law taper, produced a size
spectrum that was approximately a delta function centered at the
lowest size. This model had the same size and Gaussian profile on
the line of sight as the fractal models, and was also viewed in
projection, but there were no large-scale correlations that could
produce the appearance of OB associations and star complexes and,
hence, no statistically significant regions larger than a pixel. A
map of this model is shown at the top left of Figure
\ref{fig:fractal7}. The $\beta=3.66$ model from the lower left of
Figure \ref{fig:fractal2} is shown at the top right for
comparison.

\begin{figure}
\epsscale{1.0} \plotone{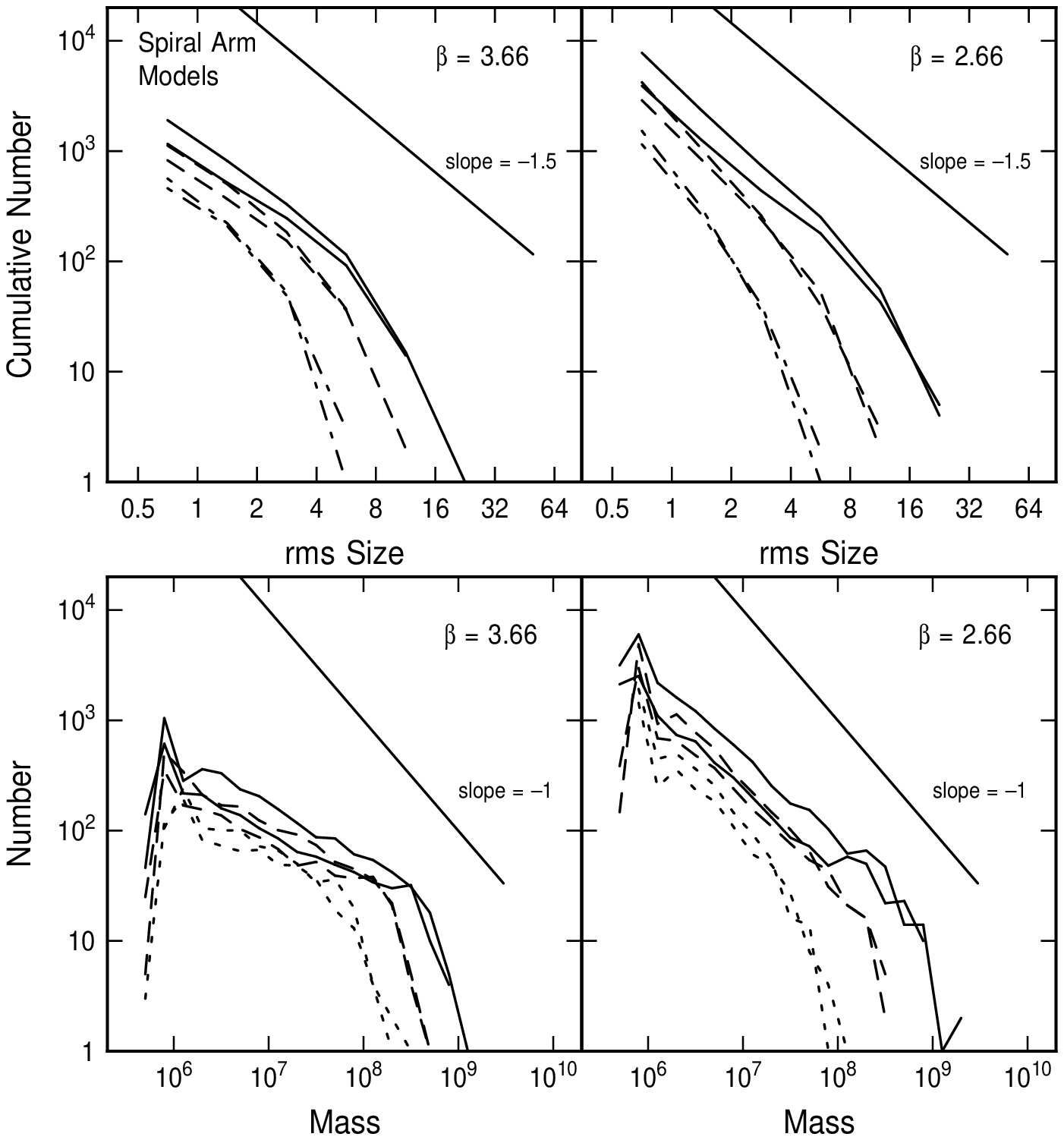} \caption{Size and mass
distribution for $\beta=3.66$ and $\beta=2.66$ models with
Gaussian profiles in one dimension of the projected plane, to
simulate spiral arms. In each panel, the solid, dashed, and dotted
line types correspond to density limits greater than 0.5 times the
peak, 0.6 times the peak, and 0.7 times the peak. The top curves
for each type correspond to broad arms with Gaussian dispersions
of 36 pixels and the bottom curves correspond to narrow arms with
dispersions of 18 pixels. The curves here are nearly the same as
in Figs. 7 and 8, except for a downward shift in each case.}
\label{fig:c12_7}\end{figure}

A third model used the same fractal algorithm as above, with the
Gaussian multiplier on the line of sight as before, but now there
was also a Gaussian multiplier along one of the projected
directions, vertical on the page. Four versions of this model are
shown in the center ($\beta=3.66$) and bottom ($\beta=2.66$) of
Figure \ref{fig:fractal7}. On the left of each, the vertical
dispersion is equal to the dispersion in depth, and on the right,
the vertical dispersion is twice the dispersion in depth. These
models are meant to represent spiral arms in the sense that the
star complexes are confined in one of the in-plane directions. The
size and mass distributions for this spiral arm model are shown in
Figure \ref{fig:c12_7}. The distributions are similar to those in
Figures \ref{fig:c1} and \ref{fig:c2} but because the areas are
smaller for the spiral models, the counts and masses of objects
are all smaller, as are the upper size and mass limits. The
implication is that the spiral arms in NGC 628 should not affect
our conclusions regarding the slope of the size, mass, and
luminosity distribution functions.  However, the minimum physical
size where the distribution function in Figure \ref{fig:c1} begins
to fall off is only 77 pc (32 px), which is more like a spiral arm
width than the galaxy image size ($\sim10$ kpc). Thus the
large-scale fall off in the spiral arm models is probably a more
reasonable explanation for the observed fall-off than the
large-scale fall off from image size limitations in Figure
\ref{fig:fractal2}.

Many other models can be imagined but they should not change the
overall size and mass distributions much as long as the density is
spatially correlated as in the fractal models with the
Kolmogorov-like values of slope for the power spectrum. For
example, the difference between star complexes with large central
clusters and star complexes without large central clusters (e.g.,
Ma\'iz-Apell\'aniz 2001), or the difference between centrally
concentrated complexes versus ring-like complexes, should not be
as nearly as great as the difference between correlated and
un-correlated models. Models in Elmegreen \& Elmegreen (2001)
using randomly placed clusters with a power law intrinsic size
distribution also failed to match the observations.

The preferred model mass distributions shown in Figure
\ref{fig:c2} are not the same as the observed luminosity
distribution except for very low $\beta$. Low $\beta$ puts
relatively more power at high spatial frequency, making relatively
more small clouds and steepening the mass function (St\"utzki et
al. 1998). This trend with $\beta$ is also apparent in Figure
\ref{massslope} for three-dimensional clouds. For realistic
$\beta$ in the range from 2.6 to 3.6, the slope of the projected
mass distribution for logarithmic intervals is between $-0.55$ and
$-0.75$ in Figure \ref{fig:c2}, clearly shallower than the
observed luminosity distribution in Figure \ref{n628lf}, where the
slope is $\sim-1$. We believe that the difference lies in the
luminosity evolution of associations and star complexes over time,
as shown by the following considerations.

Suppose the mass distribution in the model has a slope $-\mu$ for
linear intervals of mass, and the size distribution has a slope
$-1-D$ for linear intervals of size (as in Sect. \ref{size}). Then
the relation between mass $M$ and size $S$ is $M\propto S^\Delta$
where $\Delta=D/(\mu-1)$. The models suggest $D\sim1.5$ and $\mu$
is between 1.55 and 1.75, so $\Delta$ is between 2.7 and 2.
Suppose also that the average age scales with the size of the
region as $T\propto S^{\tau}$, as found in the LMC (Efremov \&
Elmegreen 1998), where $\tau\sim0.5$. This age scaling could
result from turbulence as could the whole hierarchical pattern,
provided $T\propto S/V$ (Elmegreen 2000) and the velocity
dispersion $V$ scales with $S^{0.5}$ (e.g., Larson 1981). Finally,
suppose that the luminosity of a region begins in proportion to
its mass and then decays like a power law with time: $L\propto
MT^{-\epsilon}$. From Figure 13 in Girardi et al. (1995), we
obtain $\epsilon\sim0.6$. Boutloukus \& Lamers (2003) got
$\epsilon=0.648$ for the Starburst99 models. A slightly different
value, $\epsilon=0.75$, was determined from BC01 models (Bruzual
\& Charlot 2003) by Larsen (2002).  It follows that luminosity and
mass are related by $L\propto M^{1-\tau\epsilon/\Delta}$. Equating
the distribution functions, $n(L)dL=n(M)dM$, gives the desired
result:
\begin{equation}
n(L)=n(M){{dM}\over{dL}}\propto M^{-\mu+\tau\epsilon/\Delta}
\propto L^{-\lambda}
\end{equation}
where
\begin{equation}
\lambda={{\mu-\tau\epsilon/\Delta}\over{1-\tau\epsilon/\Delta}}
\end{equation}
With the above parameters, the slope of the luminosity function,
$-\lambda$, should range between $-1.6$ and $-1.9$ for $\mu=1.55$
and $1.75$, respectively (the three values of $\epsilon$ give
about the same results). The latter $\lambda$ is reasonably close
to the observed slope in Figure \ref{n628lf}.

\section{Conclusions}

HST observations of the nearby spiral galaxy NGC 628 reveal a
regular hierarchy of structures for the brightest stars. These
structures are of a type previously studied individually as star
complexes, OB associations, OB subgroups, and so on, but in the
present study they all blend together without distinction as parts
of a nearly scale-free distribution of young stellar objects. Such
scale-free distributions have now been found using many different
techniques. The structure implies that stars form in the densest
regions of a pervasive multi-fractal gas. The most likely origin
for this gas structure is a scale-free forcing of gas motions and
compressions by turbulence and self-gravity.  The largest scale
appears, in other studies, to be approximately the ambient Jeans
length in a galaxy, suggesting that self-gravity is a primary
driver for star formation, with the substructure provided by
self-gravity, turbulence, and other forced motions.

The luminosity distribution of the stellar aggregates is
approximately a power law, like the size distribution.  The
observations show some downward curvature in both of these
distributions, suggesting functions more like log-normals than
strict power laws, but observational limitations at the small and
large ends could produce this curvature artificially. Thus we
cannot determine yet the full functional form of the
distributions.  Hierarchical structure is revealed because the
high resolution and wide field of the HST observations display a
large dynamic range for spatial scales.

Multifractal models with some resemblance to the density
structures produced by turbulence can reproduce the observed size
distribution function in NGC 628. The model mass functions are
more shallow than the observed luminosity function, but this could
be the result of luminosity fading with age if the larger regions
are older on average, in proportional to the square root of their
size.  The fractal dimension of the stellar aggregates is about
$-1.5$, similar to the fractal dimension found elsewhere for gas
distributions and other star fields.

This research was supported by NASA grant HST-GO-10402-A. D.M.E.
gratefully acknowledges the hospitality of the Space Telescope
Science Institute staff during her visit as a Caroline Herschel
Visitor in October 2005, and thanks Benne Holwerda for advice on
using $\it{SExtractor}$.

\clearpage
\begin{deluxetable}{ccc}
\tablecaption{Observations\label{tab:whole}} \tablehead{
\colhead{} & \colhead{Outer Region}&
\colhead{Central Region}\\
\colhead{RA,Dec}& \colhead{01 36 30.98, +15 48 50.23}&
\colhead{01 36 40.60, +15 46 39.00}\\
\colhead{Filter}& \colhead{Exposure (sec)}& \colhead{Exposure
(sec)} } \startdata
F435W&1200&1358\\
F555W&1000&858\\
F658N&-&1422\\
F814W&-&922\\
\enddata
\end{deluxetable}

\clearpage

\newpage
\newpage

\newpage

\newpage

\newpage

\newpage

\newpage

\newpage

\newpage

\newpage

\newpage

\newpage


\begin{thebibliography}{}
\bibitem[]{659} Bally, J., Cunningham, N., Moeckel, N., \& Smith, N.
2005, IAU Symposium 227, 12

\bibitem[]{662} Bastian, N., Gieles, M., Efremov, Yu. N., \& Lamers,
H. J. G. L. M. 2005, A\&A, 443, 79

\bibitem[]{665} Battinelli, P., Efremov, Y., \& Magnier, E.A. 1996,
A\&A, 314, 51

\bibitem[]{668} Bonnell, I. A., Bate, M. R., \& Vine, S. G. 2003,
MNRAS, 343, 413

\bibitem[]{} Boutloukos, S.G. \& Lamers, H.J.G.L.M. 2003, MNRAS, 338, 717

\bibitem[]{671} Brandeker, A., Jayawardhana, R., \& Najita, J. 2003,
AJ, 126, 2009

\bibitem[]{674} Bresolin, F., Kennicutt, R.C., Jr., Ferrarese, L.,
Gibson, B. K., Graham, J. A., Macri, L. M., Phelps, R. L., Rawson,
D. M., Sakai, S., Silbermann, N. A., and 2 coauthors 1998, AJ,
116, 119

\bibitem[]{679} Bruzual, G., \& Charlot, S. 2003, MNRAS, 344, 1000

\bibitem[]{681} Chen, Y., \& Crone, M.M. 2005, AAS, 20711305

\bibitem[]{683} Crone, M.M. 2005, AAS, 20711315

\bibitem[]{685} Dahm, S. E., \& Simon, T. 2005, AJ, 129, 829

\bibitem[]{687} Dickey, J.M., McClure-Griffiths, N.M., Stanimirovic,
S., Gaensler, B.M, \& Green, A.J, 2001, ApJ, 561, 264

\bibitem[]{690} Efremov, Y.N. 1995, AJ, 110, 2757


\bibitem[]{693} Efremov, Y.N., \& Elmegreen, B.G. 1998, MNRAS, 299,
588

\bibitem[]{707} Elmegreen, B.G. 2000, ApJ, 530, 277

\bibitem[]{696} Elmegreen, B.G., Efremov, Y.N., Pudritz, R., \&
Zinnecker, H. 2000, in Protostars and Planets IV, ed. V. G.
Mannings, A. P. Boss, \& S. S. Russell, Tucson: Univ. Arizona
Press, 179

\bibitem[]{701} Elmegreen, B.G. \& Elmegreen, D. M. 2001, AJ, 121,
1507

\bibitem[]{704} Elmegreen, B.G., Kim, S., \& Staveley-Smith, L.
2001, ApJ, 548, 749


\bibitem[]{709} Elmegreen, B.G. 2002, ApJ, 564, 773

\bibitem[]{711}  Elmegreen, B.G., Elmegreen, D.M., \& Leitner, S.N.
2003, ApJ, 590, 271

\bibitem[]{714}  Elmegreen, B.G., Leitner, S.N., Elmegreen, D.M., \&
Cuillandre, J.-C. 2003, ApJ, 593, 333

\bibitem[]{717} Elmegreen, B. C. 2004, in The Formation and
Evolution of Massive Young Star Clusters, ASP Conference Series,
Vol. 322, eds. H.J.G.L.M. Lamers, L.J. Smith, \& A. Nota,  San
Francisco: Astronomical Society of the Pacific, 277

\bibitem[]{722} Elmegreen, B.G. 2005, in The many scales in the
Universe - JENAM 2004 Astrophysics Reviews, from the Joint
European and National Astronomical Meeting,  Kluwer Academic
Publishers, ed. Jose Carlos del Toro Iniesta, et al., in press


\bibitem[]{728} Elmegreen, B.G., \& Scalo, J. 2004, ARAA, 42, 211


\bibitem[]{731} Falgarone, E., Phillips, T., \& Walker, C.K. 1991,
ApJ, 378, 186

\bibitem[]{734} Feitzinger, J. V., \& Galinski, T. 1987, A\&A, 179,
249

\bibitem[]{737} Feitzinger, J. V., \& Braunsfurth, E. 1984,
A\&A, 139, 104


\bibitem[]{742} Girardi, L., Chiosi, C., Bertelli, G., Bressan, A.
1995, A\&A, 298, 87

\bibitem[]{745} Gomez, M., Hartmann, L., Kenyon, S. J., \& Hewett,
R. 1993, AJ, 105, 1927

\bibitem[]{748}Gusev, A. S. 2002, A\&AT, 21, 75

\bibitem[]{750} Harris, J., \& Zaritsky, D. 1999, AJ, 117, 2831

\bibitem[]{752} Heydari-Malayeri, M., Charmandaris, V., Deharveng,
L., Rosa, M.R., Schaerer, D., \& Zinnecker, H. 2001, A\&A, 372,
495

\bibitem[]{756} Hodge, P.W. 1976, ApJ, 205, 728

\bibitem[]{758} Ivanov, G.R. 2005, PASRB, 4, 75

\bibitem[]{760}Ivanov, G.R., Popravko, G., Efremov, Yu.N., Tichonov,
N.A., \& Karachentsev, I.D. 1992, A\&A Supp. Ser., 96, 645

\bibitem[]{765}Kennicutt, R.C. \& Hodge, P.W. 1976, ApJ, 207, 36

\bibitem[]{767}Kennicutt, R.C. \& Hodge, P.W. 1980, ApJ, 241, 573

\bibitem[]{763}Kennicutt, R.C., Edgar, B.K., \& Hodge, P.W. 1989, ApJ, 337, 761

\bibitem[]{769}Krumholz, M.R. \& McKee, C.F. 2005, ApJ, 630, 250

\bibitem[]{771} Larson, R.B. 1981, MNRAS, 194, 809

\bibitem[]{773} Larsen, S. S. 2002, AJ, 124, 1393

\bibitem[]{775}Lelievre, M., \& Roy, J.-R. 2000, AJ, 120, 1306

\bibitem[]{777} Ma\'iz-Apell\'aniz, J. 2001, ApJ, 563, 151

\bibitem[]{779} Maragoudaki, F., Kontizas, M., Kontizas, E.,
Dapergolas, A., \& Morgan, D.H. 1998, A\&A, 338, L29

\bibitem[]{782} Nanda Kumar, M.S., Kamath, U.S., \& Davis, C.J.
2004, MNRAS, 353, 1025

\bibitem[]{785}Oey, M.S., King, N.L., \& Parker, J.W. 2004, AJ, 127,
1632

\bibitem[]{788} Oey, M.S., Watson, A.M., Kern, K., \& Walth, G.L.
2005, AJ, 129, 393

\bibitem[]{791} Pietrzyn'ski, G., Gieren, W., Fouqué, P., \& Pont,
F.A. 2001, A\&A, 371, 497

\bibitem[]{794} Sanchez, N., Alfaro, E.J., \& P\'erez, E. 2005,
astroph/0512243

\bibitem[]{797} Sharina, M.E., Karachentsev, I.D., \& Tikhonov, N.A.
1999, Astron. Letters, 25, 322

\bibitem[]{800} Smith, N., Bally, J., Shuping, R. Y., Morris, M., \&
Kassis, M. 2005, AJ, 130, 1763

\bibitem[]{803} Smith, M. D., Gredel, R., Khanzadyan, T., \&
Stankeinst, T. 2005, MmSAI, 76, 247

\bibitem[]{806} Sohn, Y.-J., Davidge, T.J. 1996, AJ, 111, 2280

\bibitem[]{808} Solomon, P.M., Rivolo, A.R., Barrett, J., \& Yahil A. 1987, ApJ, 319, 730

\bibitem[]{810} Stanimirovic, S., Staveley-Smith, L., Dickey, J.M.,
Sault, R.J., \& Snowden, S.L. 1999. MNRAS, 302, 417

\bibitem[]{813} St\"utzki, J., Bensch, F., Heithausen, A.,
Ossenkopf, V., \& Zielinsky, M. 1998, A\&A, 336, 697

\bibitem[]{816} Testi, L., Sargent, A.I., Olmi, L., \& Onello, J.S.,
2000, ApJ, 540, L53

\bibitem[]{819} Tully, R.B. 1988, Nearby Galaxies Catalog, Cambridge
Univ. Press

\bibitem[]{822} V\'azquez-Semadeni, E. 1994, ApJ, 423, 681

\bibitem[]{824} Weinberg, D. H., \& Cole, S. 1992, MNRAS, 259, 652

\bibitem[]{826} Westpfahl, D.J., Coleman, P.H., Alexander, J.,
Tongue, T. 1999, AJ, 117, 868

\bibitem[]{829} Zhang, Q., Fall, S.~M., \& Whitmore, B.~C. 2001,
ApJ, 561, 727


\end{thebibliography}
\end{document}